# Depth dependence of itinerant character in Mn-substituted $Sr_3Ru_2O_7$

G Panaccione[1,9], U Manju[2], F Offi[3], E Annese[1], I Vobornik[1], P Torelli[1], Z H Zhu[4], M A Hossain[4], L Simonelli[5], A Fondacaro[5], P Lacovig[6], A Guarino[7], Y Yoshida[8], G A Sawatzky[4] and A Damascelli[4]

[1] Istituto Officina dei Materiali CNR, TASC, in Area Science Park, S.S. 14, Km 163.5, I-34149 Trieste, Italy

[2] The Abdus Salaam International Center for Theoretical Physics (ICTP), P.O. Box 586, I-34014 Trieste, Italy

[3] CNISM and Dipartimento di Fisica Università Roma Tre, via della Vasca Navale 84, I-00146 Rome, Italy

[4] Department of Physics and Astronomy, University of British Columbia, Vancouver, British Columbia V6T,1Z1, Canada

[5] European Synchrotron Radiation Facility, B.P. 220, F-38042 Grenoble, France

[6] Sincrotrone Trieste S.C.p.A., in Area Science Park, S.S. 14, Km 163.5, I-34012 Trieste, Italy

[7] CNR-SPIN Lab. Supermat and Dipartimento di Fisica, Università di Salerno, I-84081 Salerno, Italy

[8] National Institute of Advanced Industrial Science and Technology (AIST), Tsukuba, 305-8568, Japan

E-mail: giancarlo.panaccione@elettra.trieste.it

**Abstract:** We present a core-level photoemission study of $Sr_3(Ru_{1-x}Mn_x)_2O_7$, in which we monitor the evolution of the Ru-*3d* fine structure versus Mn substitution and probing depth. In both Ru *3d*$_{3/2}$ and *3d*$_{5/2}$ core levels we observe a clear suppression of the metallic features, i.e. the screened peaks, implying a sharp transition from itinerant to localized character already at low Mn concentrations. The comparison between soft and hard x-ray photoemission, which provides tunable depth sensitivity, reveals that the degree of localized/metallic character for Ru is different at the surface than in the bulk.

[9]Author to whom any correspondence should be addressed.

**Contents**

## 1. Introduction

The change in carrier density induced by chemical doping is one of the most commonly used techniques to tailor novel properties in a variety of materials, including the strongly correlated electron systems. Transition metal oxides (TMO) are paradigmatic in this sense: they often display many different electronic phases that are quite close in energy. A comprehensive description of TMO can be reached only by disentangling such different, yet comparable, energy scales. Research on the RuO family $(Sr,Ca)_{n+1}Ru_nO_{3n+1}$ has recently suggested a conceptually different approach, based on the substitution of 4$d$ Ru with a 3$d$ transition metal impurity. A central feature in the physics of Ru-oxides is the spatial extension of 4$d$ orbitals: the properties of ruthenates are extremely sensitive to the orbital degrees of freedom, resulting in almost equal chance of displaying itinerant or localized behavior. The substitution of Ru 4$d$ with more localized 3$d$ metal atoms will strongly influence the orbital population, possibly resulting in orbital-induced novel properties. Example of this approach is the use of chromium ($Cr^{4+}$) in $SrRu_{1-x}Cr_xO_3$ and $CaRu_{1-x}Cr_xO_3$ to stabilize itinerant ferromagnetism from the normal paramagnetic state [1,2]. More recently, a 5% Ru-Mn substitution in the bilayer compound $Sr_3Ru_2O_7$ was proven to change the ground state from a paramagnetic metal to an unconventional, possibly Mott-like antiferromagnetic insulator [3,4]. The interplay between the extended, yet anisotropic Ru-4d and O-2p bonds and the localized Mn 3d-impurity states was shown to be responsible for a crystal-field level inversion, with Mn not exhibiting the expected $Mn^{4+}$ valence but rather acting as an $Mn^{3+}$ acceptor [5], a behavior bearing interesting similarities to the dilute magnetic semiconductor Mn-doped GaAs.

Unravelling what ultimately drives the electronic and magnetic properties of the $Sr_3(Ru_{1-x}Mn_x)_2O_7$ doped system is strictly linked to a direct measure of the electronic charge distribution over Ru and O-ligand orbitals. Since it is well known that electron correlations in TMO, and in particular in ruthenates, are severely influenced by the surface environment (cleavage plane, electronic/structural reconstruction, defects) [6-9], a comparison with reliable bulk-sensitive probes is mandatory. Photoemission spectroscopy (PES) possesses all the necessary characteristics in elucidating the aforementioned rich physics. In particular, core-level PES probes the different electronic screening channels via the energy location and relative intensity ratio of specific peaks, in a chemical selective way. This possibility is confirmed by recent experimental results [10], where a systematic study of doping and dimensionality effects in the core-level of various ruthenates has been carried out at fixed photon energy (Al $K\alpha$ radiation, 1486.7 eV).

In the present paper, we report a study of the Ru-$3d$ core-level fine structure vs. Mn concentration in $Sr_3(Ru_{1-x}Mn_x)_2O_7$ by soft and hard x-ray PES (HAXPES), hence with tunable depth sensitivity. The choice of focusing on the Ru-3$d$ core-levels, as opposed for instance to the Mn-2$p$ spectra, is dictated by their sharper nature which allows tracking more precisely the fine satellite structure (we measured Mn-2$p$ spectra by HAXPES, obtaining results analogous to those reported in Ref 10; no extra satellites were observed). Already at low energy, a Mn-induced suppression of the screened metallic features is observed in both Ru-$3d_{3/2}$ and $3d_{5/2}$, implying a transition from itinerant to localized character, in analogy with the reported metal-to-insulator transition [3,5]. HAXPES data confirm the change upon Mn substitution, with clear indication of a stronger electronic localization at the surface than in the bulk. Our results suggest a way to control, in the same material, metallicity of the surface-interface region vs. the bulk one, by exploiting the highly sensitive response of conducting perovskites to impurities.

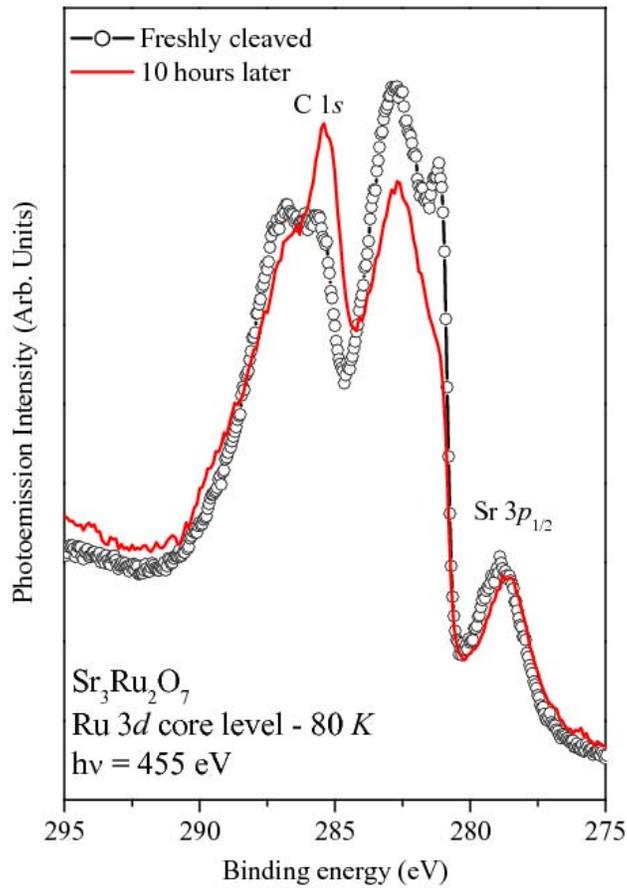

**Figure 1**. Evolution vs. time of the Ru-$3d$ core-level spectra from $Sr_3Ru_2O_7$ collected at hν=455 eV (T=80 K) on a freshly cleaved surface and after 10 hours. In the later case, the intense C-$1s$ contribution located at ~286 eV BE is clearly visible, while the screened peaks on the low BE side of both Ru-$3d_{3/2}$ and $3d_{5/2}$ lines are almost absent.

## 2. Experimental methods

High quality single crystals of $Sr_3(Ru_{1-x}Mn_x)_2O_7$, with x=0, 0.05 and 0.2 were grown by the floating zone technique. PES measurements were performed after fracturing the samples in UHV using two experimental setups: APE beamline for low-energy PES (Elettra, hν = 455 eV, base pressure $1\times10^{-10}$ mbar) [11], and VOLPE spectrometer for HAXPES (beamline ID16 at ESRF, hν = 7595 eV, base pressure $6\times10^{-10}$ mbar) [12]. The spot size in the normal emission geometry was 50x120 μm² in both cases, and the overall beamline-analyzer energy resolution was set to 200 (APE) and 350 meV (VOLPE). The Fermi energy and overall energy resolution were estimated by measuring a polycrystalline Au foil in thermal and electric contact with the samples. Identical results have been obtained consistently on several cleaved samples. The cleanliness of the surface was checked by monitoring the C-$1s$ and O-$1s$ spectra. With soft x-ray, a new cleave was needed every ~ 4 hours; instead, no traces of contamination were observed, over two days and at any temperature, in HAXPES measurements.

## 3. Results and Discussion

Surface sensitive (hν = 455 eV) Ru-$3d$ core-level spectra from $Sr_3(Ru_{1-x}Mn_x)_2O_7$ are presented in Figure 1 and Figure 2. In Figure 1 we identify the Sr-$3p_{1/2}$ peak at about 279 eV of binding energy (BE) and the Ru spin-orbit split doublet $3d_{5/2}$ and $3d_{3/2}$ in the 280-295 eV BE range. Both Ru-$3d_{5/2}$ and $3d_{3/2}$ spectra display multiple components, which have already been observed in the ruthenates; it is generally agreed that the low BE features are not induced by surface-related chemical states, and their spectral weight increases when the system enters a metallic regime[10,13—16] More specifically, each spin-orbit partner is comprised of a low BE peak, corresponding to the relaxed lowest-energy core-hole state and referred to as the screened state, and a broader higher BE structure associated with the unscreened core-hole state. The remarkable sensitivity to surface environment is observed via the evolution vs. time of the Ru-$3d$ spectral lineshape in pure $Sr_3Ru_2O_7$ . While on freshly cleaved surfaces the various spectral components are well separated, surface contamination strongly changes the lineshapes, as evidenced by: i) intense C $1s$ structures at ~285 eV BE; ii) a suppression of the screened peaks; iii) the appearance of a shoulder at ~288 eV of BE. We emphasize that the average photoelectron mean-free-path at this photon energy ranges from 4 to 8 Å [17].

In Figure 2 we present the evolution of the Ru-$3d$ spectrum as a function of Mn substitution $x$. The intense screened features, for both Ru-$3d_{5/2}$ and $3d_{3/2}$, are severely suppressed already at $x$= 5% and reduce to only a weak shoulder for $x$= 20%; an energy shift for the screened features of up to 100 meV for 20% Mn substitution is observed, as in Ref [10]. In addition, the difference spectra at the bottom of Figure 2 highlight the redistribution of spectral weight upon Mn substitution; their lineshape reveals their fine structure (with sizeable intensities around 281.3 and 285.5 eV BE), and clear shoulders on the high BE side of each spin-orbit partner (centered at 287 and 283 eV BE). These broad shoulders can be ascribed to multiplet structure which becomes more prominent upon Mn substitution, possibly suggesting the evolution from itinerant to localized character. It is interesting to note that both 5% and 20% Mn-doped spectra display close similarities with Ru-$3d$ core level results from $Ca_2RuO_4$, a pure antiferromagnetic insulator [18]. This suggests that the suppression of the screened features is compatible with metal-insulator transition induced by Mn substitution in $Sr_3(Ru_{1-x}Mn_x)_2O_7$.

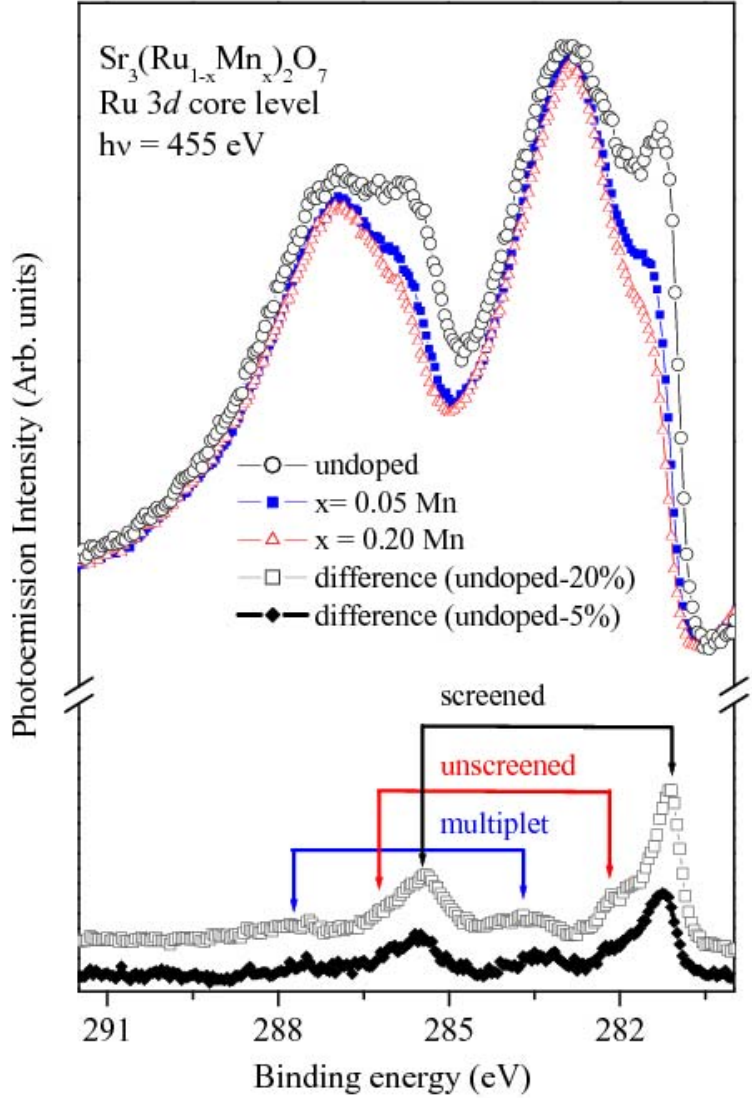

**Figure 2**. Ru-3$d$ core-level spectra collected at h$\nu$=455 eV (T=80 K) on $Sr_3(Ru_{1-x}Mn_x)_2O_7$ for x=0,0.05 and 0.2. The spectra have been normalized at the Sr-3$p_{1/2}$ peak (279 eV BE, see Figure 1). A clear suppression of the screeened features located at the low BE side of both spin-orbit partner is observed upon Mn-doping. In the bottom of the figure, the difference spectra, obtained after subtraction of the Mn-doped spectrum from the undoped one ((undoped - 20%) and (undoped-5%)) are presented, highlighting the change of spectral shape. The three bars indicate the energy position of the three sets of doublets corresponding to the main intensities (screened, unscreened and multiplet). The three doublets have been used to fit the experimental spectra in Figure 4. A shift in the energy position upon Mn doping is observed for the peaks as well as for the multiplet contribution on the high BE side.

As for the underlying driving mechanism of the transition, a purely electronic scenario was proposed based on the detection by REXS of an associated magnetic superstructure [4] and on the comparison of linear dichroism XAS data and density functional theory calculations [5]. While the evolution of the screened states in the present XPS study is compatible with the proposed electronic-driven transition induced by Mn impurities playing the role of $Mn^{3+}$ acceptors [4,5], PES does not allow determining the precise location of additional holes in the Ru-oxide host [10]. One could expect that two components should be observed in the Ru core level spectra, associated with $Ru^{4+}$ and $Ru^{5+}$ respectively. For 5 to 20% Mn substitution, at most 5 to 20% Ru atoms would be in a *5+* oxidation state, i.e. a very unfavorable case for measuring different contributions in the broad Ru-*3d* structure, where at least 80% of the intensity is of $Ru^{4+}$ character. Moreover, the induced holes would most likely be in delocalized states involving oxygen ligands, which would make their detection in XPS even less likely. One might speculate that the lack of detection of a $Ru^{5+}$ oxidation state, in ours and Ref. 10 XPS results, supports a scenario in which the Mn-induced holes are indeed delocalized.

The evolution of MIT described in Ref. 3 indicates a progressive increase of resistivity vs. Mn doping, e.g. resistivity increases a factor 10 between pure $Sr_3Ru_2O_7$ and 5% Mn-doped one. The gradual suppression of electrical conductivity upon Mn-doping, measured by a bulk property like resistivity is reflected in the decrease of the screening features in photoemission. It is important to remind, however, that photoemission, with varying degree of probing depth, is a surface sensitive technique. Surface sensitive spectra in Figure 2 display a particularly pronounced decrease of the screened intensity already at Mn 5% doping. The 5% spectrum is indeed closer to the 20% than one might expect based on the bulk properties, suggesting that MIT is more pronounced at the surface than in the bulk. The residual intensity observed in the screened features of Figure 2 upon Mn-doping can be ascribed to the gradual evolution from metallic to insulator-like character. Although the evolution of the PES intensity clearly identifies the trend, a quantitative analysis is impossible, in particular because the studied system is not a true large gap insulator.

The extreme sensitivity of the screened features -- and hence of the available screening channels -- to the local environment, calls for a more accurate determination of the metallic-insulating character upon Mn substitution. To this end, in Figure 3 we present HAXPES results which guarantee a bulk sensitivity ~8 nm [19]. The relative intensity of Sr and Ru peaks is significantly different with respect to the soft x-ray data (Figure 1 and Figure3a), which stems from the change in *p/d* cross-section ratio when passing from soft to hard x-ray PES [19, 20]. Zooming on the Ru-*3d* region (Figure 3b), screened peaks are again observed in pure $Sr_3Ru_2O_7$ together with a clear decrease of their intensity upon Mn substitution. Although the evolution as a function of Mn concentration is confirmed, the relative peak-height ratio between unscreened and screened features is reversed with respect to the surface sensitive results. Based on the enhanced probing depth of HAXPES [17,19] the intensity increase of screened peaks observed on $3d$-based TMO in the hard x-ray regime has been recently interpreted as a fingerprint of different screening mechanisms between surface and volume [21-25]. HAXPES results in vanadates across the metal-insulator transition also confirm the relationship between low BE features and metallicity [26].

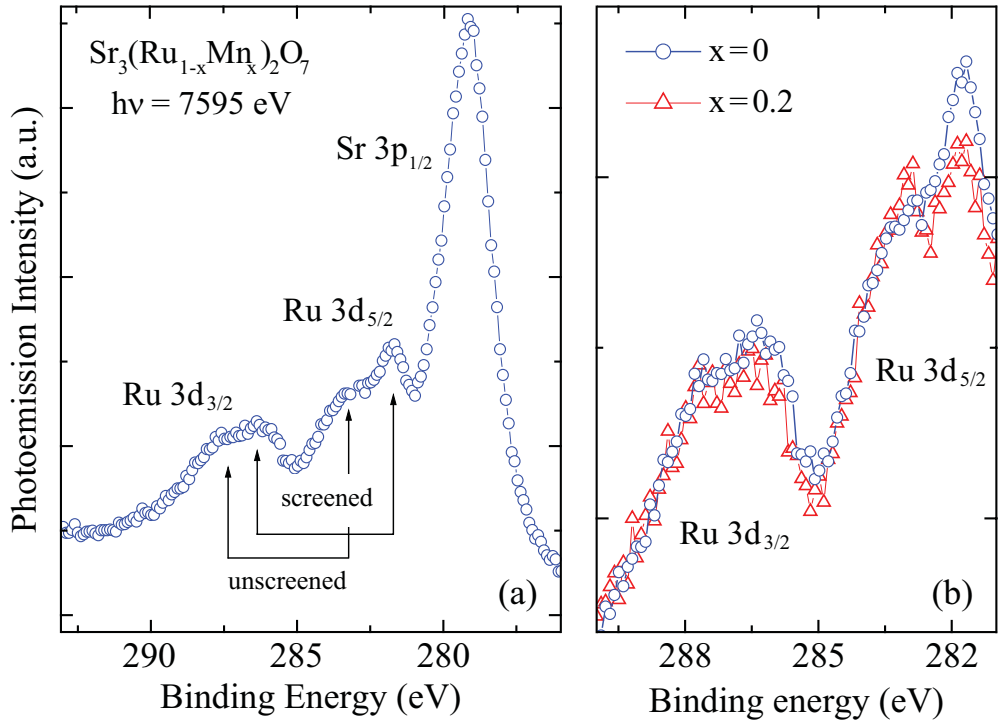

**Figure 3**. Ru-$3d$ core-level spectra acquired at at hv=7595 eV (T=20 K) on $Sr_3(Ru_{1-x}Mn_x)_2O_7$ for x=0 and 0.2. (a) Spectral region including the Sr-$3p_{1/2}$ core level; the Sr-$3p$ intensity is higher than the Ru-$3d$ one, due to the photoionization cross section increase in the HAXPES regime. (b) Enlarged view of the Ru-$3d$ energy range; for clarity, the spectra have been normalized to the Sr-$3p_{1/2}$ intensity. A reversed peak-height ratio between screened and unscreened features, for both $3d_{5/2}$ and $3d_{3/2}$, is observed.

We stress that our data cannot provide direct evidence for the exclusive electronic-driven picture and we have only a compatible scenario with the one described in Ref.4 by Hossain *et al.*. While we can directly show by soft and hard x-ray photoemission, and the relative intensity of screened peaks, that the surface is more correlated than the bulk, we cannot address directly the role and interplay of other degrees of freedom, such as charge, spin, and lattice.

To better disentangle screened and poorly screened features and also to identify any additional components, we performed a spectral decomposition by fitting routines. Each peak was represented by a symmetric function generated by a Lorentzian lineshape convoluted with a Gaussian. The Lorentzian function represents the lifetime broadening effect, while the Gaussian accounts for all other broadenings including energy resolution (350 meV for HAXPES and 200 meV for soft x-ray). The results of these spectral decompositions are presented in Figure 4, with the corresponding parameters summarized in Table 1.

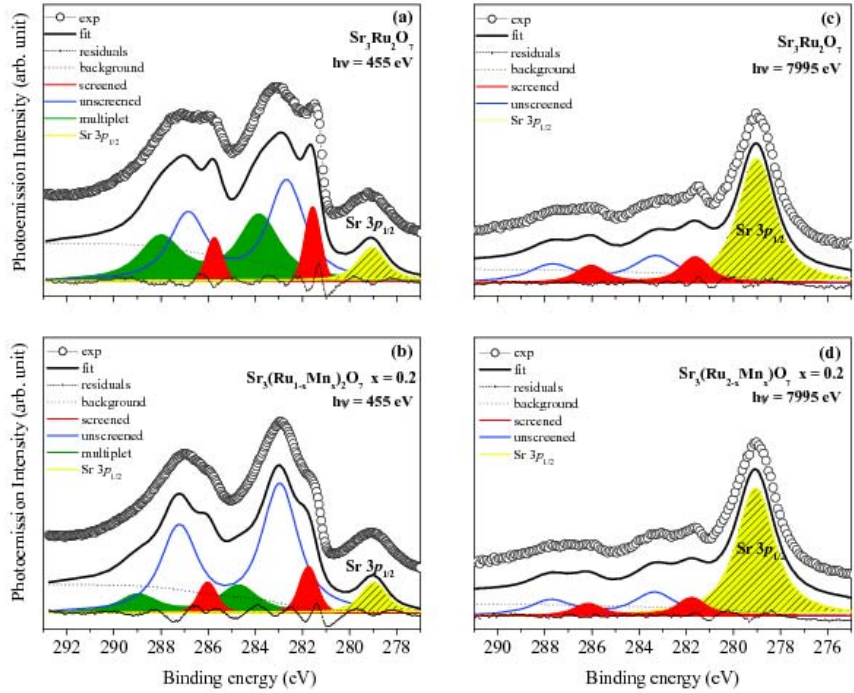

**Figure 4**. Fit of the Ru-3$d$ spectra measured at hν = 455 (a,b) and 7595 eV (c,d). The Gaussian instrumental resolution broadening is ~350 meV at 7595 eV and ~200 meV at 455 eV; a background proportional to the integrated intensity and residual intensity are also included (black dotted curve and dotted curve, respectively). For the Ru-3$d$ doublets (blue curves, red curves and green curves) a ~4 eV spin-orbit splitting and a 1:1.5 statistical intensity ratio are imposed. Energy position of peaks are normalized to the Sr 3p$_{1/2}$ peak (yellow curves)

The open circles show the experimental spectra and the solid line shows the fitting decomposition. The spectra recorded at 455 eV photon energy could be best fit using such function with three sets of doublets (screened, unscreened and multiplet peaks), where the intensity ratio between the spin-orbit split components has been fixed at 1.5, as determined by the degeneracy ratio. The multiplet structures appearing on the higher BE side, for both 3$d_{5/2}$ and 3$d_{3/2}$ components, have been modeled with Gaussian intensities and constrained to the energy position of the structures appearing in Figure 2. The fitting procedure confirms the reduction of such multiplet contribution upon Mn substitution. For the bulk sensitive spectra recorded at hν = 7595 eV, the best fit is obtained with two sets of doublets only; in this case adding a multiplet contribution does not increase significantly the quality of the fit. This is partly due to the residual intensity at high BE arising from the Sr-3$p$ core-level, which limits the analysis of fine details in the HAXPES spectra. The integrated intensity ratio between screened and unscreened peaks is summarized in Table 1. At each photon energy, the values are identical within the error bars, for both 3$d_{5/2}$ and 3$d_{3/2}$ components, confirming that: i) no extra intensity (hence a different ratio for different spin-orbit partners) arising from contamination is found, knowing that C 1$s$ signal overlaps with 3$d_{3/2}$ ; ii) the evolution of the metallic/insulating behavior upon Mn-substitution is different for bulk and surface; iii) as for the surface sensitive data, the results of the fit indicate not only a shift to higher BE of the screened peaks but also a shift, in the same direction, of the unscreened contributions. All these findings point again to an important modification of the surface electronic structure compared to the bulk one.

Table 1 Parameters from the Ru-3$d$ core-level spectral analysis. $I_S/I_U$ refers to the integrated intensity ratio of the Ru-3$d_{5/2}$ screened (S) and unscreened (U) peaks; within uncertainties, identical values are found for the Ru-3$d_{3/2}$ features.

| hν (eV) | Compound | $I_S/I_U$ |
|---|---|---|
| 455 | $Sr_3Ru_2O_7$ | 0.23 |
| 455 | $Sr_3(Ru_{1.6}Mn_{0.4})O_7$ | 0.16 |
| 7595 | $Sr_3Ru_2O_7$ | 0.89 |
| 7595 | $Sr_3(Ru_{1.6}Mn_{0.4})O_7$ | 0.63 |

The presence and evolution of screened and unscreened features in the *4d* PES spectra from ruthenates have been described in terms of a Mott-Hubbard picture within a DMFT approach [27]; cluster calculations of Ru-3$d$ in $Sr_2RuO_4$ suggest that the energy separation between screened and unscreened peaks is reminiscent of the Coulomb interaction between Ru-3$d$ and 4$d$ holes, and comparable with the Ru 4$d$-$t_{2g}$ bandwidth *W* [28]. In particular, smaller $U_{dd}$ values correspond to higher screened intensity; thus, the observed difference between surface and bulk sensitive PES spectra suggests a stronger localization in the surface and subsurface region, as possibly due to the reduced coordination or surface relaxation. The interpretation in terms of final-state screening properties, supported by both experimental and theoretical considerations, suggests also that the linewidth of the screened peak bears a degree of proportionality to the Ru bandwidth *W* [27-29]; due to the increase of electron localization in passing from volume to surface, one would expect a line narrowing in surface sensitive PES, reflecting the progressive reduction of *W*. This argument appears to be corroborated by the linewidth fit results for the screened peak in HAXPES and soft x-ray PES. Although the experimental energy resolution is similar for the different kinetic energy ranges, the Lorentzian contribution of the screened features is sharper in the soft x-ray (FWHM ~300 meV) than in the HAXPES regime (FWHM~600 meV). Note, however, that a conclusive unambiguous statement on this latter point is prevented by the broad nature of the experimental features, which makes the value of this observation purely qualitative.

### 4. Conclusions

In conclusion, the $Sr_3(Ru_{1-x}Mn_x)_2O_7$ metal-insulator transition has been studied by variable probing depth PES. The measured evolution of the Ru 3$d$ core-level signals, provides evidence for a progressive increase of electron correlations, not only upon doping but also from bulk to surface. The reduction of metallic –like screening channels, which is more pronounced in the vicinity of the surface might also play a role at interfaces, leading to potentially new physical properties.

**Acknowledgements**

We acknowledge fruitful discussions with M.A. van Veenendaal, M. Grioni and A. Vecchione. This work was supported by INFM-CNR, the Killam Program (A.D.), the Alfred P. Sloan Foundation (A.D.), CRC Program (A.D., G.A), NSERC, CFI, CIFAR Quantum Materials, and BCSI.